\documentclass[twocolumn,amsmath,amssymb]{revtex4}
\usepackage{graphicx}

\begin{document}

\title{Detecting competing orders through the edge states in the heterostructures with high-Tc superconductors}

\author{Tao Zhou$^{1}$}
\email{tzhou@scnu.edu.cn}

\author{Yi Gao$^{2}$}

\author{Z. D. Wang$^{3}$}
\email{zwang@hku.hk}

\affiliation{
$^{1}$Guangdong Provincial Key Laboratory of Quantum Engineering and Quantum Materials,
and School of Physics and Telecommunication Engineering,
South China Normal University, Guangzhou 510006, China\\
$^{2}$ Department of Physics and Institute of Theoretical Physics,
Nanjing Normal University, Nanjing, 210023, China.\\
$^{3}$Department of Physics and Center of Theoretical and Computational Physics, The University of Hong Kong, Pokfulam Road, Hong Kong, China.
}

\date{\today}
\begin{abstract}
Considering a $d$-wave superconductor being in proximity to a two-dimensional Weyl model, a topological superconductor with gapless edge states may be realized. We here demonstrate that the system can become topologically trivial when an additional d-density-wave order is also included. The edge states become gapped and may be detected in experiments. For this, the d-density-wave order can be detected experimentally, so that different scenarios for the pseudogap in high-T$_c$ superconductors may be distinguished. Moreover,
this method may be used to detect various hidden orders through putting high-T$_c$ superconductors being in proximity to various topological nontrivial materials.

\end{abstract}
\maketitle


Although the high-T$_c$ superconductivity was discovered in cuprate materials more than thirty years ago~\cite{bed},  no consensus has so far been reached regarding to its mechanism.
This is partially due to lack of a profound understanding of the pseudogap state~\cite{tim,mue,eme,cha,kei,frad,vis}. Generally there are two different scenarios for the pseudogap state, according to
the relationship between the pseudogap and the superconducting pairing gap. One is the phase fluctuation scenario, suggesting that the pseudogap is due to the preformed Cooper pairs~\cite{eme}.
The other scenario lies in that the pseudogap may be due to certain competing hidden order~\cite{cha,kei,frad,vis}. Actually,
the existence of possible competing orders in high-T$_c$ superconductors is a rather important issue. Now it was widely believed that multiple competing orders may exist and they are not limited to explain the pseudogap phenomenon~\cite{kei,frad,vis,dod,cai}.
Identifying various orders in the superconducting state is important and may find a useful clue in searching for the real origin of superconductivity.

Recently, research on topological superconductors has also attracted tremendous interest~\cite{xlqi}. A topological superconductor is characterized by a full superconducting gap in the
bulk and topologically protected gapless states at the system edges. The edge states are in conjunction
with the Majorana bound states, which obey non-Abelian statistics and have potential applications in topological quantum
computations~\cite{naya}. Notably, most previous efforts have been made in searching for topological superconductors and Majorana bound states.
It was proposed theoretically that an effective topological superconductor may be realized in the heterostructure system, e.g., the superconductor is in proximity to a topological insulator~\cite{lfu}, quantum anomalous insulator~\cite{qi}, or a semiconductor with the spin-orbital coupling~\cite{sau,lutc,oreg}.
Experimentally, the above heterostructure systems were indeed realized and
some signatures of Majorana bound states were reported in these systems~\cite{wil,rok,den,das,mou,pxu,sun,he}. On the other hand, since the high-T$_c$ superconductors have much larger pairing gap, it is natural to consider whether the topological superconductor can be realized in the high-T$_c$ superconductor families. So far, the cuprate-based heterostructure has been studied intensively~\cite{zare,wan,yil,zxli,yan},
and several signatures of proximity induced superconductivity were indeed probed experimentally~\cite{zare,wan}.
Very recently, it was also indicated that the iron-based superconductors may provide another platform  based on high-T$_c$ superconductors for realizing
the topological superconductors~\cite{zhao,zhang,wang}.

In this paper, we elucidate that the topological heterostructure may provide a useful platform to detect and resolve the possible competing orders in high-T$_c$ superconductors.
As is known, in the superconducting state of high-T$_c$ superconductors, it may be rather difficult to detect the possible competing orders because such order is normally covered up by the superconducting gap.
Even in the case that some signatures of certain order are seen, it is still difficult to determine its physical origin because the energy spectrum may be qualitatively the same for different theoretical scenarios.
While in the topological system, there exist gapless edge states, such that
 even if the bulk states are similar to each other for different orders, the edge states can be significantly different. Thus we may resolve different competing orders through studying the edge states. Especially, the topological protected features are usually sensitive to the Hamiltonian symmetries, which may be exploited to determine the physical origin of the pseudogap behavior and
 to probe some weak hidden orders.

To demonstrate our proposal, we start from a two-dimensional lattice-Weyl model, given by,
\begin{equation}
H_N=\sum\limits_{{\bf k}\sigma}\varepsilon_{{\bf k}\sigma} c^\dagger_{{\bf k}\sigma}c_{{\bf k}\sigma}+
\sum\limits_{\bf k}(\lambda_{\bf k}c^\dagger_{{\bf k}\uparrow}c_{{\bf k}\downarrow}+h.c.),
\end{equation}
with $\varepsilon_{{\bf k}\sigma}=-2\sigma t(\cos k_x+\cos k_y)$, being a spin polarized hopping term. $\lambda_{\bf k}=2\lambda_0(\sin k_x+i\sin k_y)$, corresponding to a spin-orbital coupling. Then we can obtain two energy bands, with $E(k_x,k_y)=\sqrt{\varepsilon_{{\bf k}\sigma}^2+\mid\lambda_{\bf k}\mid^2}$. These two bands as functions of the momentums are plotted in Fig.~1. The Weyl points at the positions $(\pm \pi,0)$ and $(0,\pm \pi)$ are seen clearly.
The above Hamiltionian may be realized in the HgTe/CdTe quantum well system~\cite{liu,ahn}, or the single layer LaCl/LaBr materials~\cite{nie}.

  \begin{figure}
\centering
  \includegraphics[width=2.4in]{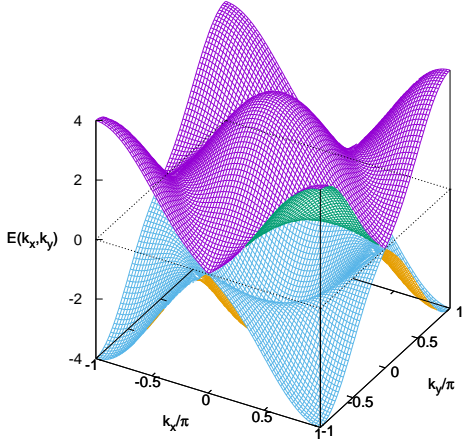}
\caption{(Color online) The normal state energy bands from Hamiltonian of Eq.~(1).
}
\end{figure}

In proximity to a cuprate high-T$_c$ superconductor, the $d$-wave superconducting (DSC) pairing term is induced to the system.
We also consider a competing d-density wave (DDW) order, which was proposed
to describe the pseudogap state in the underdoped high-Tc cuprates~\cite{cha}.

The Hamiltionian of the DSC and DDW parts are expressed as,
\begin{eqnarray}
H_{DSC}=\sum_{\bf k}\Delta_{\bf k}(c^\dagger_{{\bf k}\uparrow}c^\dagger_{{\bf -k}\downarrow}+h.c.),
\end{eqnarray}
and
\begin{equation}
H_{DDW}=\sum_{\bf k\sigma}W_{\bf k} (c^{\dagger}_{{\bf k}\sigma}c_{{\bf k}+{\bf Q}\sigma}).
\end{equation}
Here ${\bf Q}=(\pi,\pi)$, $\Delta_{\bf k}=2\Delta_{DSC}(\cos k_x-\cos k_y)$ and $W_{\bf k}=2i\Delta_{DDW}(\cos k_x-\cos k_y)$.

We define the topological invariant $\mathcal{N}$, express as,
\begin{equation}
\mathcal{N} = \frac{1}{2\pi}\iint[\frac{\partial{a_{y}({\bf k})}}{\partial{k_{x}}}-\frac{\partial{a_{x}({\bf k})}}{\partial{k_{y}}}]dk_x dk_y,
\end{equation}
with
\begin{equation}
a_{\alpha}{(\bf k)}=-i\sum\limits_{m\in occ}\langle{{u_m}({\bf k})\mid\frac{\partial}{\partial{k_{\alpha}}}\mid u_m ({\bf k})}\rangle,
\end{equation}
where $\mid u_m ({\bf k})\rangle$ is the eigenstate of the occupied state $m$.

To study the edge states, we define a partial
Fourier transformation along the $x$-direction with $C^\dagger_{{\bf k}}=\frac{1}{{\sqrt {N_xa}}}\sum_x C^\dagger_{{k_y}}(x)e^{ik_x x}$.
The Hamiltonian is reduced to the quasi-one-dimensional one, which can be rewritten as,
\begin{eqnarray}
H_N=&-t\sum\limits_{k_y,x,\sigma}[\sigma c^\dagger_{k_y\sigma}(x)c_{k_y\sigma}(x+a)+h.c.]\nonumber\\&-i\lambda_0\sum\limits_{k_y,x}[c^\dagger_{k_y\uparrow}(x)c_{k_y\downarrow}(x+a)+h.c.]\nonumber\\
\nonumber\\&-2t\sum\limits_{k_y,x,\sigma}
\sigma\cos k_yc^\dagger_{k_y\sigma}(x)c_{k_y\sigma}(x)\nonumber\\&+2i\lambda_0\sum\limits_{k_y,x}\sin k_yc^\dagger_{k_y\uparrow}(x)c_{k_y\downarrow}(x),
\end{eqnarray}
with $a$ being the lattice constant along the $x$-direction.

The Hamiltonian for the DSC pairing and DDW parts are rewritten as,
\begin{eqnarray}
H_{DSC}=&\sum\limits_{k_yx}[\Delta_{DSC}c^\dagger_{{k_y\uparrow}}(x)c^\dagger_{{-k_y\downarrow}}(x\pm a)+h.c.]\nonumber\\&-\sum\limits_{k_yx}[2\Delta_{DSC}\cos k_yc^\dagger_{{k_y\uparrow}}(x)c^\dagger_{{-k_y\downarrow}}(x)+h.c.]
\end{eqnarray}
and
\begin{eqnarray}
H_{DDW}=&-\sum\limits_{k_yx\sigma}[i\Delta_{DDW} c^\dagger_{{k_y\sigma}}(x)c_{{k_y+\pi\sigma}}(x+a)+h.c.]\nonumber\\&-\sum\limits_{k_yx\sigma}[2i\Delta_{DDW}\cos k_y  c^\dagger_{{k_y\sigma}}(x)c_{{k_y+\pi\sigma}}(x)].
\end{eqnarray}

The whole hamiltonian can be written as $4N_x\times 4N_x$ or $8N_x\times 8N_x$ (with the DDW order) matrix form. Then the $x$-dependent spectral functions $A_x(k_y,\omega)$ and the Local density of states $\rho_x(\omega)$
are express as,
\begin{equation}
A_{{x}}({k_y},\omega)=\sum_{n,\sigma} \frac{\mid u^{n}_{{x}\sigma}({k_y})\mid^2}{\omega-E_n({k_y})+i\Gamma},
\end{equation}
and
\begin{equation}
\rho_{x} (\omega)=\sum_{k_y}A_{x}({k_y},\omega).
\end{equation}
 $u^{n}_{{x}\sigma}$ and $E_n({\bf k})$ are eigenvectors and eigenvalues through diagonalizing the Hamiltonian matrix.

In the following, 
the input parameters are chosen as $t=1$, $\lambda_0=0.5$, $a=1$, and $\Gamma=0.01$. We have checked numerically that our main results are not sensitive to the parameters.

   \begin{figure}
\centering
  \includegraphics[width=3.5in]{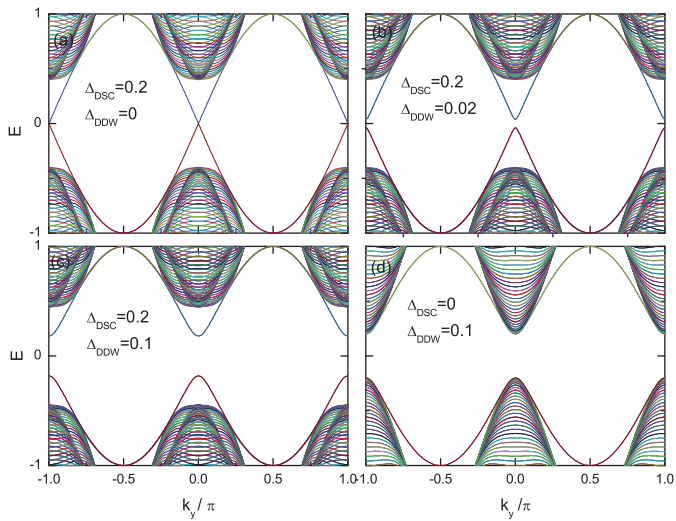}
\caption{(Color online) Energy eigenvalues of the Hamiltonian in different states (a pure dsc state (a), the coexistence states (b \& c), and a pure DDW state (d) ), with the open boundary condition along $x$-direction being considered.
}
\end{figure}

We first study the energy spectra of different states, including the pure DSC state, the DDW state, and the state in which the DSC order and the DDW order coexist. Note that, the bulk energy bands from all of the three states are fully gapped and a topological invariant [Eq.(4)] is well defined. We now study numerically the edge states of these three states through considering the open boundary condition along the $x$-direction.
Then the numerical results of the energy bands for the reduced quasi-one-dimensional Hamiltonian are presented in Fig.~2. In the pure DSC state with $\Delta_{DDW}=0$, as is seen in Fig.~2(a), there are in-gap edge states crossing the Fermi energy at the momentum $k_y=0$ and $k_y=\pi$. The edge states connect the upper and lower energy bands, indicating that the system may be a topological superconductor.
The energy bands for the coexisting state are displayed in Figs.2(b) and 2(c). For this case, the energy bands at the system edges are also fully gapped, with the gap magnitude depending on the DDW intensity. For the energy band of the pure DDW state, as presented in Fig.~2(d), it is also fully gapped for both system bulk and system edge.
Thus for both the coexisting state and the pure DDW state, the system turns to be topological trivial. This result may be used to detect the DDW order experimentally.
The numerical results for the energy spectra are consistent with the numerical calculations of the topological invariant. For the pure DSC state, the topological invariant $\mathcal{N}$ equals to $2$ obtained from Eq.(4), corresponding to the two unequivalent Weyl points at the momentums $(0,\pi)$ and $(\pi,0)$ shown in Fig.~1.
When the DDW term is added, the topological invariant immediately turns to zero, as a result, the energy bands at the system edge are also fully gapped.

The above results can be understood based on a symmetry analysis.
Here the topological non-trivial behavior is related to the band inversion.
We define a typical spin and momentum inversion operator $\mathcal{P}$, with
\begin{equation}
\mathcal{P} {\bf k}\mathcal{P}^{-1}=-{\bf k},\qquad \mathcal{P} c_{\uparrow(\downarrow)}{\mathcal{P}}^{-1}=c_{\downarrow(\uparrow)}.
\end{equation}
Then we have the anticommutation for the normal state band, with $\{H_N,\mathcal{P}\}=0$.
This anticommutation relation is corresponding to the band inversion when both the spin and the momentum are reversed.
With the superconducting pairing term, this band inversion symmetry preserves. Actually, here the topological behavior is stabilized by this symmetry (named as $\mathcal{P}$ symmetry in the following). While the DDW Hamiltonian does not have $\mathcal{P}$ symmetry, as a result, the system becomes topological trivial even when the DDW magnitude is rather small.

Experimentally, the possible edge states may be detected by the angle-resolved photoemission spectroscopy (ARPES) experiments~\cite{dama} or the scanning tunneling microscopy (STM) techniques~\cite{oys}. The results of these two experiments can be described theoretically by the spectral function [Eq.(9)] and the LDOS [Eq.(10)], respectively. We first study numerically the spectral function considering the open boundary condition along the $x$-direction with $N_x=100$. The intensity plots of the spectral functions for the three different states at the system bulk and the system edge are presented in Fig.~(3). At the system bulk with $x=50$ [Figs.~3(a)-3(c)], the spectra are fully gapped for all of the three states we considered.
It seems that there is no significant difference between the bulk spectra of different states.
Especially, when the DDW order coexists with the DSC order, the spectrum is almost the same as that of the pure DSC state, as is seen in Fig.~3(a) and 3(b). Thus here the DDW gap is indeed covered up and seems rather difficult to be detected merely from the bulk spectra.

 \begin{figure}
\centering
  \includegraphics[width=3.5in]{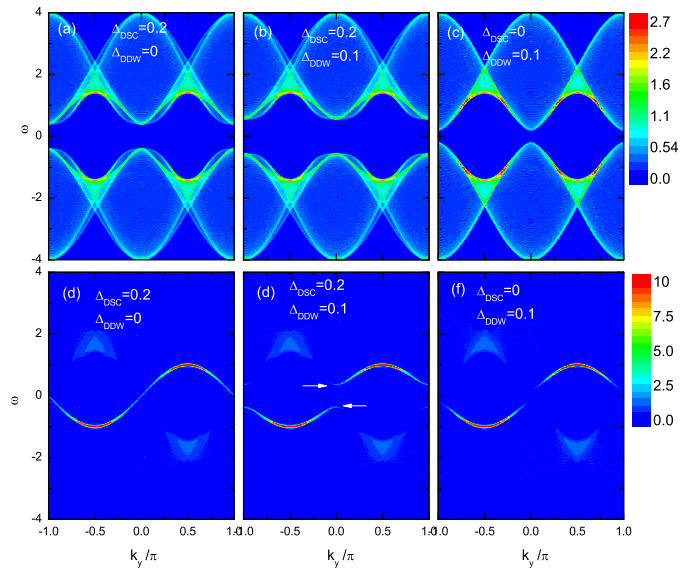}
\caption{(Color online) The intensity plots of spectral functions at the system bulk (a-c) and the system edge (d-f) for different states.
}
\end{figure}

 We now turn to address the edge states of these three states. The numerical results for the spectral functions at the system edge $(x=1)$ are presented in Figs.~3(d)-3(f). In the pure DSC state [Fig.~3(d)], there are gapless edge states crossing the Fermi energy. While when a DDW term is added [Fig.~3(e)], the edge states end up at a nonzero finite energy, indicated by the arrows.
An obvious energy gap is seen clearly. In the pure DDW state [Fig.~3(f)], an obvious energy gap is also seen clearly. Therefore, the existence of the DDW order may indeed be detected from the edge states by ARPES experiments.

 Now we study the LDOS spectra. Considering the open boundary along the $x$ direction, the LDOS from the system bulk to the system edge are plotted in Fig.~4. First let us look at the spectra in the system bulk. The intensities reach zero value at low energies for all of the three states, indicating the fully gapped feature. Also, in the coexisting state, the DDW gap is almost hidden by the DSC gap and may be difficult to be detected experimentally.

Let us study the LDOS spectra at the system edge. In the pure DSC state, as is seen in Fig.~4(a),
the LDOS intensities at low energies are nonzero, due to the gapless edge states. While when an additional DDW order is added to the system [Fig.~4(b)], the low energy intensities recover to the zero value. Especially, there exist two obvious low energy peaks lying symmetric at the two sides of the Fermi energy. These two low energy peaks are corresponding to the energy gap opened by the DDW term at the system edge, as indicated by arrows in Fig.~3(e). This additional gap feature can be seen clearly from the LDOS spectrum in the coexisting state. In the pure DDW state [Fig.~4(c)], the LDOS spectrum is fully gapped and no in-gap features exist. We conclude that different states may be resolved well from the LDOS spectra at the system edge and detected by STM experiments.

   \begin{figure}
\centering
  \includegraphics[width=3.5in]{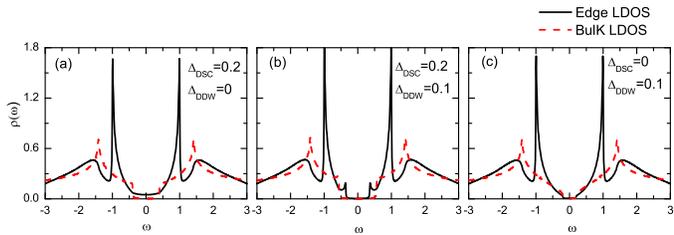}
\caption{(Color online) The LDOS for different states.
}
\end{figure}

We have demonstrated that the topological property varies when a DDW order is added, then the existence of the DDW order may be detected through investigating the edge states.
On the other hand, the phase fluctuation scenario involves no symmetry breaking and the topological feature are usually not sensitive to the fluctuated phase. This is significant different from the DDW scenario. Thus our present proposal may also be used to
distinguish different pseudogap scenarios.

Our main results are understood well based on the symmetry analysis.
Based on our present work, some useful symmetric information about the possible hidden order may be provided.
It is insightful to investigate the coexistence of the superconducting order with some other possible orders which preserve the $\mathcal{P}$ symmetry, and compare the numerical results with those for DDW order. We now consider two kinds of spin order, namely, ferromagnetic (FM) order and antiferromagnetic (AFM) order, with the corresponding Hamiltonian being expressed as,
\begin{equation}
H_{FM}=\sum\limits_{{\bf k}\sigma}\sigma \Delta_{FM}c^\dagger_{{\bf k}\sigma}c_{{\bf k}\sigma},
\end{equation}
and
\begin{equation}
H_{AFM}=\sum\limits_{{\bf k}\sigma}\sigma \Delta_{AFM}(c^\dagger_{{\bf k}\sigma}c_{{\bf k}+{\bf Q}\sigma}+h.c.).
\end{equation}
Now we discuss whether the topological behavior is still stable when the FM or AFM order coexists with the DSC pairing order.
Considering the open boundary condition, the energy bands of the two kinds of coexisting states are plotted in Fig.~5.
As is seen, for both cases, there are gapless edge states at $k_y=0$ and $k_y=\pi$. The numerical results for the energy bands are consistent with the calculations of the topological invariant, namely, we have $\mathcal{N}=2$ for both coexisting states.
Thus the topological nontrivial behavior is robust and stable when the FM or AFM order is added into the system. This is also consistent with the symmetry analysis, namely, these two spin orders do not break the $\mathcal{P}$ symmetry.

\begin{figure}
\centering
\includegraphics[width=3.5in]{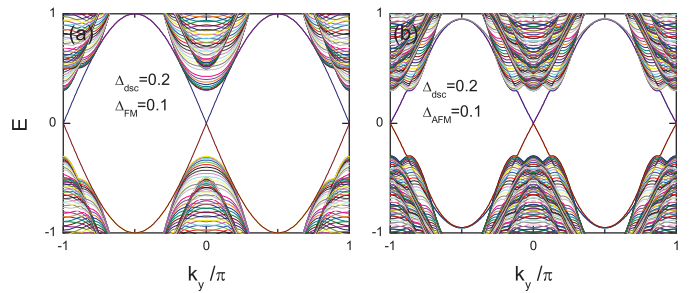}
\caption{(Color online) Energy eigenvalues of the Hamiltonian when the superconducting order coexists with the spin order, with the open boundary condition along $x$-direction being considered.
}
\end{figure}

 We here would like to remark the significance of the present work. Firstly, we have provided an effective method to detect possible weak hidden orders in high-T$_c$ superconductors. Some useful information may be provided based on the symmetry analysis. And our results may be used to
 differentiate different pictures of the pseudogap behavior. Secondly, we here emphasize that our starting minimal model is considered just for illustration. One may also choose other topological nontrivial system to detect other possible competing orders, e.g, if a topological insulator system with time reversal symmetry is considered, then a topological superconductor protected by the time reversal symmetry may be constructed. This system may be used to detect both the spin order and the DDW order because the time-reversal symmetry is broken by these orders.
Moreover, we expect that our scenario may also work well for iron-based superconducting materials.
Recently it has been reported the topological superconductor can be realized in the family of iron-based superconductors. And it was widely believed that multiple competing orders may also be important in understanding the superconductivity of this family.
At last, we here provided an effective method to realize the topological superconductor with a high-T$_c$ superconductor platform. The results may be useful in the further studies of the topological superconductors and Majorana bound states and may have potential application in topological quantum computation.

We thank Wei Chen for helpful discussions. 
This work was supported by the GRF of Hong Kong (Grant Nos. HKU173309/16P and HKU173057/17P), the Natural Science Foundation from Jiangsu Province of China (Grant No. BK20160094), and the Start-up Foundation from South China Normal University.

\end{document}